\documentclass[prl,a4paper,superscriptaddress,twocolumn,amsmath,amssymb,floatfix,amstex]{revtex4-1}%

\newcommand{\ktimes}{\rangle\! \langle}
\newcommand{\op}[2]{|#1\ktimes #2|}

\usepackage{graphicx,graphics,rotating}	
\usepackage{dcolumn}
\usepackage{bm}
\usepackage{hyperref}
\usepackage{color}
 \usepackage{soul}   								
\newcommand{\new}[1]{#1}

\usepackage{aurical}
\usepackage[T1]{fontenc}
\newcommand{\raur}{\text{\Fontauri r}}

\begin{document}
\title{Two \new{critical} localization lengths in the Anderson transition on random graphs}

\author{I.~Garc\'ia-Mata}
\affiliation{Instituto de Investigaciones F\'isicas de Mar del Plata
(IFIMAR), CONICET--UNMdP,
Funes 3350, B7602AYL
Mar del Plata, Argentina.}
\affiliation{Consejo Nacional de Investigaciones Cient\'ificas y
Tecnol\'ogicas (CONICET), Argentina}
\author{J.~Martin}
\affiliation{Institut de Physique Nucl\'eaire, Atomique et de
Spectroscopie, CESAM, Universit\'e de Li\`ege, B\^at.\ B15, B - 4000
Li\`ege, Belgium}
\author{R.~Dubertrand}
\affiliation{Institut f\"ur Theoretische Physik, Universit\"at Regensburg, 93040 Regensburg, Germany}
\author{O.~Giraud}
\affiliation{LPTMS, CNRS, Univ.~Paris-Sud, Universit\'e Paris-Saclay, 91405 Orsay, France}
\author{B.~Georgeot}
\affiliation{%
Laboratoire de Physique Th\'eorique, IRSAMC, Universit\'e de Toulouse, CNRS, UPS, France
}
\author{G.~Lemari\'e}
\email[Corresponding author: ]{lemarie@irsamc.ups-tlse.fr}
\affiliation{%
Laboratoire de Physique Th\'eorique, IRSAMC, Universit\'e de Toulouse, CNRS, UPS, France
}

\date{\today}
\begin{abstract}
We present a full description of the nonergodic properties of wavefunctions on random graphs without boundary in the localized and critical regimes of the Anderson transition. We find that they are characterized by two \new{critical} localization lengths: the largest one describes localization along rare branches and diverges \new{with a critical exponent $\nu_\parallel=1$} at the transition. The second length, which describes localization along typical branches, \new{reaches at the transition a finite universal value (which depends only on the connectivity of the graph), with a singularity controlled by a new critical exponent $\nu_\perp=1/2$}. We show numerically that these two localization lengths control \new{the finite-size scaling properties} of key observables: wavefunction moments, correlation functions and spectral statistics. Our results are \new{identical} to the theoretical predictions for the typical localization length in the many-body localization transition, with the same critical exponent. This strongly suggests that the two transitions are in the same universality class and that our techniques could be directly applied in this context.
\end{abstract}

\maketitle

\textit{Introduction.} There has been a huge interest recently in the nonergodic properties of many-body states \cite{PhysRevLett.78.2803, PhysRevB.91.081103, De_Luca_2013, Pino536, doi:10.1002/andp.201600284, doi:10.1002/andp.201600350, PhysRevB.96.214205, 10.21468/SciPostPhys.4.6.038,  Monthus_2016, mace2018multifractal, luitz2014universal, PhysRevE.86.021104, PhysRevB.95.195161, PhysRevB.99.104206, SciPostPhys.2.2.011, PhysRevLett.122.106603}, in particular related to many-body localization (MBL) \cite{basko2006metal, PhysRevLett.95.206603} (see \cite{nandkishore2015many, abanin2017recent, alet2018many, abanin2018ergodicity} for recent reviews). In such problems, the structure of Hilbert space is tree-like, and it was found that in many cases the states do not explore ergodically all the branches. The problem of Anderson localization on random graphs is a simple one-particle model which is believed to capture this physics, and has attracted recently a strong interest \cite{abou1973selfconsistent, castellani1986upper,mirlin1994distribution,monthus2008anderson,biroli2012difference,deluca2014anderson,monthus2011anderson,facoetti2016non,sonner2017multifractality,parisi2018anderson,kravtsov2018non,savitz2019anderson,tikhonov2016fractality,biroli2018delocalization,tikhonov2016anderson,PhysRevLett.118.166801, tikhonov2019statistics,tikhonov2019critical,kravtsov2015random,altshuler2016nonergodic,monthus2016localization, de2019survival, bera2018return, PhysRevB.96.201114, PhysRevB.96.064202, PhysRevE.90.052109, PhysRevB.95.104206, PhysRevLett.117.104101, pino2019ergodic}. On the finite Bethe lattice (tree with boundary) there is now a consensus that there is a transition from a localized to a nonergodic delocalized phase \cite{monthus2011anderson, tikhonov2016fractality, sonner2017multifractality, parisi2018anderson, kravtsov2018non, biroli2018delocalization, facoetti2016non, savitz2019anderson}. For generic random graphs (with loops and without boundary) the situation is still debated but several numerical and analytical studies point toward a transition from a localized to an ergodic delocalized phase with however nonergodic properties below a certain scale which diverges exponentially at the transition \cite{tikhonov2016fractality,biroli2018delocalization,tikhonov2016anderson, PhysRevLett.118.166801, tikhonov2019statistics,tikhonov2019critical}. However, a precise description of this nonergodic behavior is still lacking.

\begin{figure*}
\includegraphics[width=0.9\linewidth]{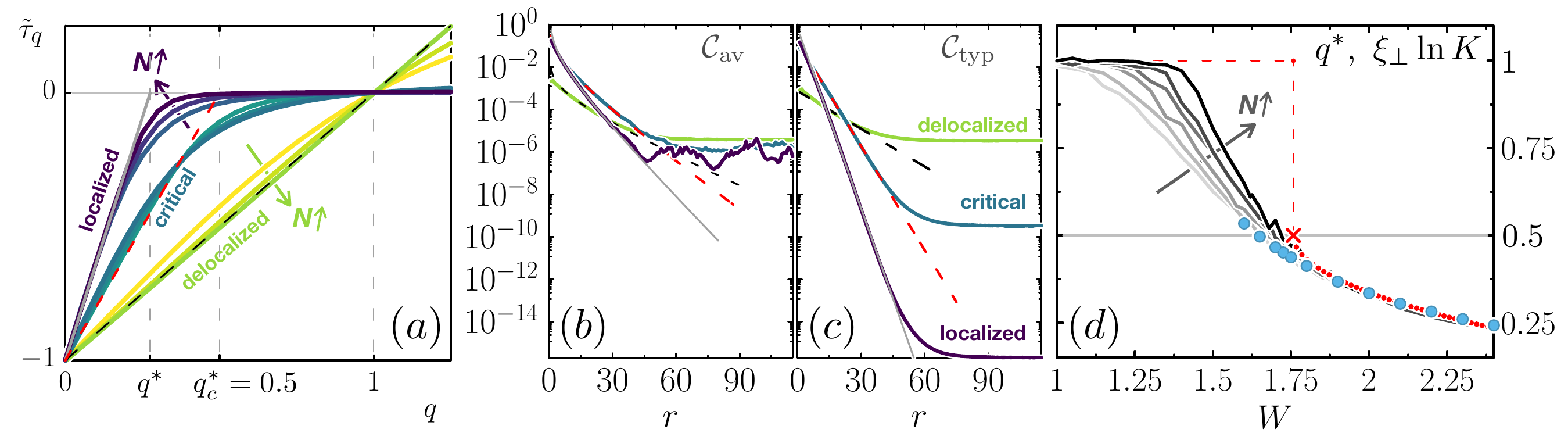}
\caption{\label{fig:tauq} (\textit{a}): \textit{Strong multifractality (resp ergodicity) of wavefunctions in the localized and critical (resp delocalized) regimes.} $\tilde{\tau}_q $ (calculated as $\log_2 \langle P_q(N/2) \rangle -\log_2 \langle P_q (N) \rangle$) vs $q$ for $W=1.05$ (delocalized), $1.725$ (critical), $2.3$ (localized) for $p=0.06$,
$N=2^{12},\, 2^{15},\, 2^{20}$; straight lines for small $q$ are fits by $q/q^*-1$. $\langle P_q \rangle$ for small $q$ is averaged, as is usual, over boxes of size $4$ along the 1D chain \eqref{model}. (\textit{b}) and (\textit{c}): \textit{Determination of $\xi_\parallel$ and $\xi_\perp$ through correlation functions $\mathcal{C}_{\text{av}}$ and  $\mathcal{C}_{\text{typ}}$} for $N=2^{19}$. Fitting lines are \new{Eq.~\eqref{eq:correl}}. (\textit{d}): \textit{Strong multifractal properties are controlled
by $\xi_\perp$.} $q^*$ (continuous lines, $N$ between $2^{15}$ and $2^{20}$) and $\xi_\perp \ln K$ (circles) vs $W$ for $p=0.06$. 
In the localized and critical regimes, \mbox{$q^* \approx \xi_\perp \ln K$}. 
Red dotted line is a fit with the critical behavior 
 $q^* = q_c^* - C (W-W_c)^{\nu_\perp}$, with \mbox{$q_c^*=0.5$}, $\nu_\perp=0.5$ and two fitting parameters $C$ and $W_c$. $W_c \approx 1.74$ agrees with other more controlled determinations (see \cite{PhysRevLett.118.166801}). Dashed line is the asymptotic ergodic behavior $q^*=1$, 
 when  $N\rightarrow \infty$ for $W<W_c$.}
\end{figure*}

At strong disorder, there are theoretical arguments (forward scattering approximation) \cite{abou1973selfconsistent, miller1994weak, somoza2007universal, monthus2009statistics} that relate Anderson localization to the problem of directed polymers. Directed polymers on trees display a glassy phase with strong non-ergodic properties, exploring only few branches instead of the exponentially many available \cite{derrida1988polymers, cook1989directed, FIM:PRB10}. The analogy thus suggests that Anderson localized states on random graphs are located on rare branches, along which they are exponentially localized with a localization length $\xi_\parallel$. \new{Up to now, $\xi_\parallel$ has been believed to be the only critical localization length, diverging with a critical exponent $\nu_\parallel =1$ at the transition \cite{zirnbauer1986localization, Mirlin_1991, PhysRevLett.67.2049, refId0, mirlin1994distribution, PhysRevLett.118.166801}.} On the other hand, recent rigorous results on the Bethe lattice \cite{aizenman2011absence, PhysRevLett.106.136804} (see also \cite{mirlin1994distribution, efetov1999supersymmetry, kravtsov2018non}) put forward another length scale $\xi_\perp$, which corresponds to the typical exponential decay of wavefunctions (excluding rare events). \new{In \cite{aizenman2011absence, PhysRevLett.106.136804}, a sufficient condition for delocalization was proven: when $\xi_\perp>\xi_\perp^c$,  $\xi_\perp^c$ being a specific finite value which depends only on the connectivity of the tree, an avalanche process occurs, where the exponential decay does not compensate anymore for the exponential proliferation of sites at distance $r$. However, the transition was still described as controlled by $\xi_\parallel$, so that the importance of  $\xi_\perp$ as a critical length was unclear, all the more since it remains finite at the transition.}

In this paper, we show that this length scale $\xi_\perp$, that we identify as governing the exponential decay away from the rare branches, \new{actually} controls important aspects of the critical behavior of key observables for generic random graphs. First, the wavefunction  moments $P_q = \sum_i \vert \psi_i \vert^{2q}$ for large 
$q$ (i.e.~$q>1$) focus on large amplitudes 
and therefore reflect the localization on the rare branches, governed by $\xi_\parallel$. In contrast, values of $q<0.5$ focus on small 
amplitudes, and reflect the bulk localization properties controlled by $\xi_\perp$.  Second, the standard average correlation function \cite{evers2008anderson, tikhonov2019statistics} is dominated by rare branches and thus by $\xi_\parallel$ whereas a suitably defined typical correlation function is controlled by $\xi_\perp$. Lastly, the behavior of spectral statistics at small energy distance is dominated by bulk localization and thus by $\xi_\perp$. 

A crucial point of our findings is that $\xi_\parallel$ and $\xi_\perp$ are associated to two different critical exponents $\nu_\parallel\approx 1$ and $\nu_\perp\approx 0.5$ which control the finite-size scaling properties of the different observables close to the Anderson transition. In particular, we show that ${\xi_\perp}^{-1} \approx  {\xi_\perp^c}^{-1} +  \xi^{-1}$, with $\xi \sim (W-W_c)^{-\nu_\perp}$. Recent theoretical results on MBL point towards very similar behavior \cite{PhysRevLett.121.140601, thiery2017microscopically, PhysRevB.99.094205}, with the MBL transition governed by a \new{similar} avalanche process when the typical decay of matrix elements reaches a universal critical value. \new{An identical} equation describes the approach to the transition, with the same critical exponent. \new{Up to now, MBL has always been believed to be somewhat similar to Anderson localization in random graphs \cite{PhysRevLett.78.2803, basko2006metal, PhysRevLett.95.206603}. Our results make this analogy more precise and} strongly suggest that the two transitions are in the same universality class. \new{This indicates that the MBL transition could be clarified with our techniques}.

\textit{Model.} We consider a generic class of random graphs \cite{ZhuXiong00, ZhuXiong01, Gir05} built by taking a one-dimensional Anderson model of $N$ sites with periodic boundary conditions, and adding  $\lfloor pN \rfloor$ shortcut links between random pairs of sites ($\lfloor pN \rfloor$ is the integer part of $pN$). The Hamiltonian reads 
\begin{equation}\label{model}
H=\sum_{i=1}^N\varepsilon_i\op{i}{i} +  \op{i}{i+1}+\sum_{k=1}^{\lfloor pN \rfloor} \op{i_k}{j_k} + h. c.
\end{equation}
The on-site disorder is described by random variables $\varepsilon_i$ of zero mean with a Gaussian distribution of standard deviation $W$. The second term describes nearest-neighbor hopping and the third term the long-range links between $(i_k,j_k)$ with $|i_k-j_k|>1$. Such a graph has locally a tree-like structure with an average branching number $K\approx 1+2p$ and an average branching distance $\approx 1/(2p)$. This type of graph 
("small world networks" \cite{watts1998collective}) 
is similar to random regular graphs when $p \rightarrow 1/2$. For all $p$, it displays loops of typical size $\sim \log N$, hence has no boundary, and the diameter (maximal distance between sites) $d_N$ increases as $\sim \log N$, making the system effectively infinite-dimensional. 
For our numerical investigations we use exact diagonalization of very large sparse matrices of sizes up to $N=2^{21}$ with the Jacobi-Davidson method \cite{JacobiDavidson, schenk2008large}, to extract 16 eigenvalues and eigenfunctions closest to $E=0$. We average over 3000 to 15000 disordered graphs depending on $N$ and denote this by $\langle \rangle$. 

This model \eqref{model} presents an Anderson transition at a certain value of disorder $W_c (p)$ \cite{Gir05}. Recently, we investigated its critical properties through finite-size scaling of wavefunction moments $P_q$ for large $q>1$~\cite{PhysRevLett.118.166801}. On the localized side these moments are controlled by  $\xi_\parallel$ which was found to diverge at the transition as $\xi_\parallel \sim (W-W_c)^{-\nu_\parallel}$ with $\nu_\parallel \approx 1$. On the delocalized side, an ergodic behavior at a scale larger than a nonergodic volume $\Lambda $ diverging exponentially at the transition as $\log \Lambda \sim (W_c-W)^{-\kappa}$ with $\kappa \approx 0.5$ was found. These observations agree with the analytical predictions of \cite{Mirlin_1991, PhysRevLett.67.2049, refId0}.

These results concern the localization length $\xi_\parallel$. We will now show that key physical observables, such as multifractality of wavefunctions, are controlled by $\xi_\perp$, and highlight its importance for the critical behavior.

\textit{Simple model for $\xi_\perp$.} 
For the finite Bethe lattice, a strong multifractal behavior \cite{castellani1986upper, evers2008anderson} was predicted to occur at the root in the localized phase \cite{tikhonov2016fractality,deluca2014anderson}, that is, for large system sizes, moments scale as $P_q \sim N^{-\tau^*_q} $ with 
\begin{equation}\label{tauqstar}
\tau^*_q= \left( \frac{q}{q^*} -1 \right) \text{ for } q< q^{*}; \;\;\; \tau^*_q= 0  \text{ for } q > q^{*}\;.      
\end{equation}
One has $q^{*}=q_c^{*}=0.5$ at the transition \cite{evers2008anderson} and $q^*$ decreases with increasing $W$ away from $W_c$ \cite{tikhonov2016fractality}. 

This behavior can be interpreted as a manifestation of $\xi_\perp$. Indeed, 
let us consider a wavefunction exponentially localized at the root of a tree with connectivity $K$ and depth $d$, with the same localization length $\xi_\perp$ along all the branches. 
For \mbox{$\xi_\perp < 1/\ln K$}, the moments
\begin{eqnarray}\label{toy}
 P_q  = \dfrac{\sum_{r=0}^{d-1} K^r [e^{-r/\xi_\perp}]^q}{\left[\sum_{r=0}^{d-1} K^r e^{-r/\xi_\perp}\right]^q} 
 \sim N^{-\tau_q^*} \; ,
 \end{eqnarray}
 where $\tau_q^*$ is given by \eqref{tauqstar}, \mbox{$q^*= \xi_\perp \ln K < 1$} and $N = K^{d}$.
Here the strong multifractal behavior \eqref{tauqstar}  is due to the exponential proliferation of sites at distance $r$ which compensates, for $q<q^*$, the exponential decrease of the localized wavefunction.  As $q^{*}=q_c^{*}=0.5$ at the transition, this 
model also suggests that the critical behavior is localized with $\xi_\perp^c = q_c^{*}/\ln K$ for the Bethe lattice. 

\textit{Moments.}
For our model \eqref{model}, we show in Fig.~\ref{fig:tauq} (left) the local $\tilde{\tau}_q = -\frac{d\log{\langle P_q\rangle}}{d\log N}$ in different regimes, localized, critical ($W=W_c$) and delocalized. In the localized regime, $\tilde{\tau}_q$ clearly tends when $N \rightarrow \infty$ towards $\tau^*_q$ in \eqref{tauqstar} with a $q^* < 0.5$. $q^*$ can be determined by a linear fit of $\tilde{\tau}_q$ at small $q$ where finite-size effects are negligible. In the critical regime the same behavior is observed, with $q^* \approx 0.5=q_c^{*}$. In the delocalized regime, the behavior clearly tends to the ergodic limit $\tau_q =q-1$ for large $N$. Indeed, our determination of $q^*$ at small $q$ gives 1 in the delocalized phase (see Fig.~\ref{fig:tauq}).

 \textit{Correlation functions.} In Fig.~\ref{fig:tauq} (middle) we show the average $\mathcal{C}_{\text{av}} (r)= \langle \sum_{i=1}^{N} \vert \psi_i \vert^2 \vert  \psi_{i+r} \vert^2 \rangle $ and typical \mbox{$\mathcal{C}_{\text{typ}} (r) = \exp \langle \ln (\sum_i\vert \psi_i \vert^2 \vert  \psi_{i+r} \vert^2) \rangle $} correlation functions, calculated along the 1D lattice, which can be seen as a typical branch. In the localized phase, we find \new{
 \begin{equation}\label{eq:correl}
  \mathcal{C}_{\text{av}}  \sim r^{-\alpha} K^{-r} \exp (-r/\xi_\parallel)   \; ; \quad \mathcal{C}_{\text{typ}}  \sim \exp (-r/\xi_\perp) \; .
 \end{equation}}
$\mathcal{C}_{\text{typ}} $ gives the typical exponential decay with $\xi_\perp$ along an arbitrarily
chosen branch, namely the 1D lattice, whereas $\mathcal{C}_{\text{av}}$\new{, which agrees with \cite{tikhonov2019statistics},} is dominated by the configurations where the rare populated
branches coincide with the 1D lattice, and is thus controlled by $\xi_\parallel$ (see \cite{morone2014large, PhysRevE.97.012152} for similar large deviations in random magnets). At criticality $\xi_\perp$ remains finite while $\xi_\parallel$ diverges, hence $\mathcal{C}_{\text{typ}} \ll \mathcal{C}_{\text{av}}$ for $W\ge W_c$. In the delocalized regime, $\mathcal{C}_{\text{av}} \approx \mathcal{C}_{\text{typ}}$, confirming ergodicity.

In Fig.~\ref{fig:tauq} (right) we show the variation of $q^*$ and $\xi_\perp$ as a function of $W$ for different system sizes $N$. In the delocalized regime $W< W_c$, $q^*$ tends towards a plateau at $q^*=1$ (dashed line), confirming ergodicity, with strong finite-size effects close to the transition. In the localized regime $W>W_c$, the data for $q^*$ and $\xi_\perp \ln K$ agree very well with each other \footnote{Here, $K$ has been determined numerically through the exponential increase of the number $N_r$ of sites at distance $r \ll d_N$, $N_r \sim K^r$.}. Thus, as in the simple model \eqref{toy}, we have $q^*=\xi_\perp \ln K$.
These results confirm that the strong multifractal properties of wavefunctions are controlled by  $\xi_\perp$, as well as the typical correlation function. Moreover, $q^*$ and $\xi_\perp \ln K$  are well-fitted by $q^*_c- C (W-W_c)^{\nu_\perp}$ (red dotted line), with $q^*_c=0.5$, $\nu_\perp=0.5$ and only two fitting parameters $C$ and $W_c$, a critical behavior that we will derive below.

At this stage, it might seem that the non-diverging value of $\xi_\perp^c$ implies a weak influence of this length on the transition. However, we will now show that this is not the case and that actually the finite-size scaling properties close to the transition of the key observables discussed above are controlled by the new critical exponent $\nu_\perp$.

\textit{Finite size scaling.} 
To address the critical properties of the transition, we use a finite-size scaling analysis \cite{fisher1972scaling}.
This is well understood in finite dimension \cite{pichard1981finite, mackinnon1981one, slevin1999corrections, rodriguez2011multifractal}, but requires some care in random graphs, due to the exponential growth of the volume with linear size. 
In \cite{PhysRevLett.118.166801}, we showed that surprisingly the moments $P_q$ for large values of $q$ follow a different scaling depending on the side of the transition. On the localized side, $P_q= P_q^c  F_\text{lin} (d_N/\xi_{\parallel} )$ with $d_N = \log N$ and $P_q^c \equiv P_q(W_c)$, indicating a linear scaling, whereas on the delocalized side $P_q= P_q^c  F_\text{vol} (N/\Lambda )$ indicating an unusual volumic scaling. This volumic scaling, together with the observed ergodic behavior at small $W$, implies an ergodic behavior for $N \gg \Lambda$ in the entire delocalized phase $W<W_c$. In infinite dimension the two types of scaling are distinct as was also observed recently for the MBL transition \cite{mace2018multifractal}.

To describe the finite-size scaling of the $P_q$ for small $q < 0.5$, we generalize the scaling assumptions of \cite{PhysRevLett.118.166801} and assume a two-parameter scaling function:
\begin{equation}\label{eq:scahypV2}
 \frac{P_q}{P_q^c} = F\left(X, Y\right), \ \quad \ X=\frac{d_N}{\xi},\, Y=\frac{N}{\Lambda},
\end{equation}
with $\xi$ and $\Lambda$ two scaling parameters.
\begin{figure}
\includegraphics[width=0.9\linewidth]{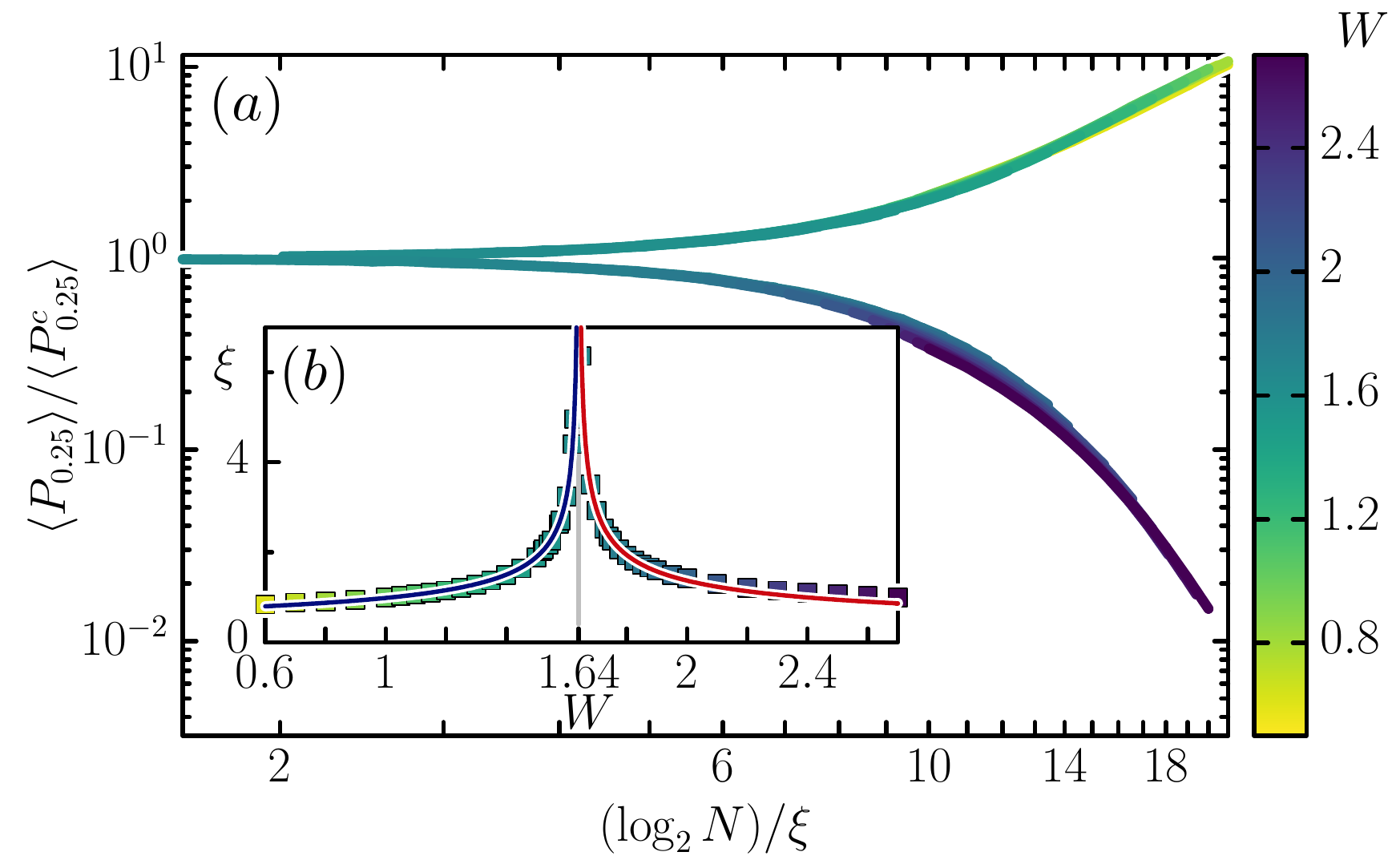}
\caption{\label{fig:scaPq025} (\textit{a}) \textit{Finite-size scaling of wavefunction moments $\langle P_q \rangle$ for small $q=0.25<0.5$}. Inset (\textit{b}) is $\xi$ vs $W$ with the fit \mbox{$\xi \sim \vert W-1.64\vert^{-\nu_\perp}$} (solid line), $\nu_\perp \approx 0.43$. $p=0.06$ and $N$ varies between $2^9$ and $2^{20}$.}
\end{figure}
%
We can then recover the observed large size behavior given by \eqref{tauqstar} in the localized phase $W>W_c$
if we further assume that $F$ in \eqref{eq:scahypV2}  has the asymptotic behavior
\begin{equation}\label{eq:assymptscaFloca}
 F(X,Y) \sim V(X)^{-Aq} + Y^{\tau_q^c}; \ W>W_c;\ X,Y\gg 1,
\end{equation}
with $ A$ a positive constant and $V(X) \sim e^X$ the volume associated with the length $X$.  Since $P_q^c \sim N^{-\tau_q^c}$, with $\tau_q^c$ given by \eqref{tauqstar} for $q^{*}=q_c^{*}$, then \mbox{$P_q^c \;\;V(X)^{-Aq}\sim N^{1-q/q^*}$}  with
\begin{equation}\label{qstarloc}
 q^* \approx \left (\frac{1}{q_c^{*}} + \frac{A}{\xi} \right)^{-1} \; \text{or} \; \; \frac{1}{\xi_\perp} \approx \frac{1}{\xi_\perp^c} +\frac{ A\ln K}{\xi} \; .
\end{equation}
Together, \eqref{eq:scahypV2} and \eqref{eq:assymptscaFloca} give $
 P_q \sim N^{1-q/q^*}  + \Lambda^{-\tau_q^c}$.
Thus \mbox{$P_q \sim N^{-\tau_q^*}$}, as observed in Fig.~\ref{fig:tauq}, which justifies the asymptotic form \eqref{eq:assymptscaFloca}.

In the delocalized regime, an ergodic behavior $P_q \sim N^{-(q-1)}$ was found for $N \gg \Lambda$ at large $q$ \cite{PhysRevLett.118.166801}, implying ergodicity at \textit{all} $q$. However, the nonergodic volume $\Lambda$ is exponentially large close to the transition, which leaves room to a nonergodic multifractal behavior at intermediate scales and thus a linear scaling ($F(X,Y)\sim$ function of $X$) associated with it.

The finite-size scaling of the $\langle P_q \rangle$ for $q=0.25$ and $p=0.06$ displayed in Fig.~\ref{fig:scaPq025} shows a very good collapse with a linear scaling on both sides of the transition. The asymptotic behavior of the scaling function is well-fitted by $F(X,Y)\sim V(X)^{-Aq}$ (see \eqref{eq:assymptscaFloca}) with $A \approx 1.9$. The scaling parameter $\xi$ is shown in the inset to diverge at the transition as $\xi \sim |W-W_c|^{-\nu_\perp}$, with $\nu_\perp \approx \tfrac{1}{2}$ on  both sides of the transition. This is in striking contrast with the result for large $q > 1$ where the localization length $\xi_\parallel$ diverges with an exponent $\nu_\parallel \approx 1$ at the transition \cite{ PhysRevLett.118.166801}. From \eqref{qstarloc} we infer that 
$q^* = q^*_c - C (W-W_c)^{\nu_\perp}$ with $C$ a constant, in perfect agreement with numerical data (see Fig.~\ref{fig:tauq}).
These results describe explicitly how $\xi_\perp$ approaches the finite $\xi_\perp^c$ close to the transition and are compatible \cite{tikhonov2019private} with the results in \cite{mirlin1994distribution, sonner2017multifractality}.

\begin{figure}
 \includegraphics[width=\linewidth]{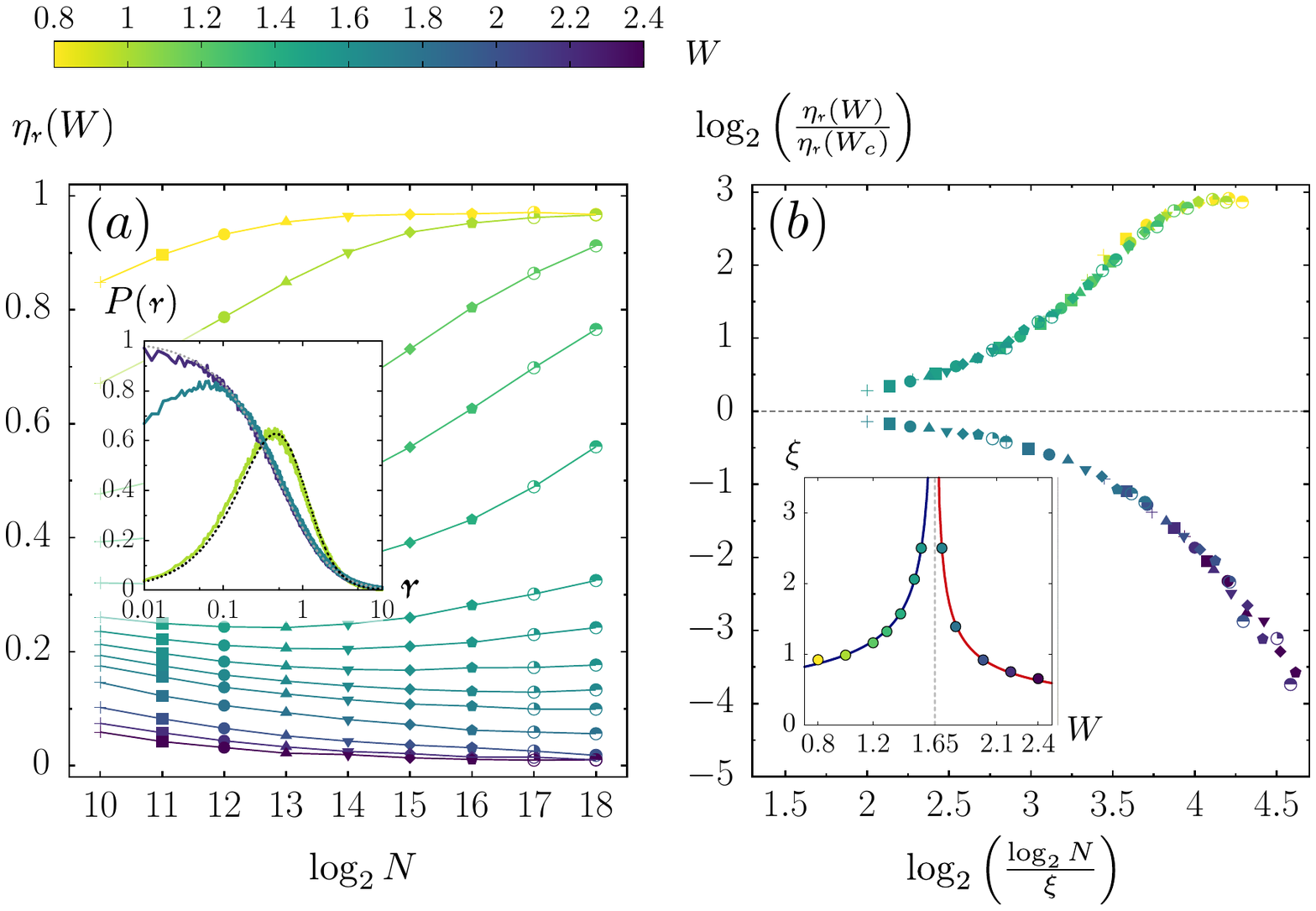}
 \caption{\label{fig:spectrum} \textit{Spectrum: finite-size scaling analysis of $\eta_\raur$} (see text) for $p=0.06$. $N=2^{10}$ to $N=2^{18}$. Each symbol is a different size. (\textit{a}): Raw data for $\eta_\raur$; inset: example of $P(\raur)$ (see text) for localized, critical and delocalized phases. (\textit{b}):
 Collapse of the data after a rescaling of the form $\eta_\raur(W)/\eta_\raur(W_c)=F_\mathrm{lin}(\log_2N/\xi)$ with $\xi$ the scaling parameter; inset: $\xi$ vs $W$ across the transition at $W_c=1.65$. Solid lines are $\xi \sim |W-1.65|^{-0.49}$ (delocalized branch) and $\xi \sim |W-1.65|^{-0.51}$ (localized branch).}
\end{figure}

\textit{Spectral statistics.} The length scale $\xi_\perp$ can also be probed using 
the distribution of the ratios $\raur$ of spacings ~\cite{PhysRevB.75.155111,PhysRevB.97.220201,PhysRevLett.110.084101} between consecutive energy levels. The transition manifests itself through a change from Poisson statistics in the localized phase to a random matrix distribution in the delocalized phase, as seen in Fig.~\ref{fig:spectrum} left. We define the parameter $\eta_\raur$ as 
$\eta_\raur=\tfrac{\langle\min\left(\raur,1/\raur\right)\rangle-I_{\mathrm{P}}}{ I_{\mathrm{WD}}-I_{\mathrm{P}}} $ 
where 
$I_{\mathrm{P}} = \langle \min\left(\raur,1/\raur\right) \rangle \approx 0.386$ (resp. $I_{\mathrm{WD}}\approx 0.536$) for Poisson statistics (resp. random matrix statistics). At the transition spectral statistics is expected \cite{tikhonov2019statistics} to converge to Poisson logarithmically slowly, which is compatible with our numerical data.

Figure~\ref{fig:spectrum} right shows the result of a finite-size scaling analysis of $\eta_\raur$ for different system sizes $N$ and disorder strengths $W$ for $p=0.06$. The raw data shown in the left panel are found to collapse after a rescaling of the form $\eta_\raur(W)/\eta_\raur(W_c)=F_\mathrm{lin}(\log_2N/\xi)$ with \mbox{$\xi=A\,|W-W_c|^{-\nu_\perp}$} and $\nu_\perp \approx 0.5$ for both sides of the transition. We note that similar scaling laws where reported in \cite{PhysRevE.72.066123} for the Bethe lattice and scale-free networks. These results indicate that the behavior at small energy distance (level repulsion) is dominated by $\xi_\perp$, the localization length associated with $\nu_\perp$. Indeed, wavefunctions at different but closeby energies are located on different branches and their overlap is controlled by $\xi_\perp$. 

\textit{Universality.} 
We have checked \cite{papierlong}
that our results are valid for $p$ up to $p=0.49$, which corresponds to random graphs with $K=1.98$, and other types of disorder distributions. We have also checked that the new critical exponent $\nu_\perp$ does not vary significantly as a function of $q<0.5$ and $0<p<0.5$. 

The above analysis is the standard procedure for finite-size scaling near a second-order phase transition, where the algebraic divergence of the scaling parameter $\xi$ gives the critical exponent of the transition. Technically, the scaling quantity is the observable divided by its behavior at criticality, for example $\langle P_q\rangle/\langle P_q^c\rangle$. Then, doing so amounts to subtract $1/\xi_\perp^c$ from $1/\xi_\perp$, and one gets a diverging length scale $\xi$. Thus $\nu_\perp$ appears as the new critical exponent of the Anderson transition for the key observables considered.

\textit{Conclusion.} Our results clearly show that there exist two different localization lengths in the Anderson transition on random graphs, $\xi_\parallel$ describing rare branches and $\xi_\perp$ describing the bulk, which control the critical behavior of different physical observables and are associated with distinct critical exponents $\nu_\parallel\approx 1$ and $\nu_\perp\approx 0.5$. 
\new{On the delocalized side, only one critical exponent $\kappa \approx \nu_\perp \approx 0.5$ controls the critical behavior of all observables considered.}
This clarifies the nature of the Anderson transition in the limit of infinite dimensionality, which remains, for the bulk properties, a continuous, second-order phase transition, while rare events, characteristic of random graphs, are responsible for the discontinuous properties described up to now. We further note that in finite dimension, the localization length critical exponent tends to $0.5$ in the limit of large dimensionality \cite{tarquini2017critical}, suggesting that the rare branch mechanism may be absent in large finite dimension (see also \cite{efetov1999supersymmetry}). \new{An interesting perspective is to investigate with these techniques the non-ergodic delocalized phase which has been demonstrated in other models \cite{kravtsov2015random, tikhonov2016fractality, kravtsov2018non, pino2019ergodic} and where another critical exponent $1$ has been found \cite{kravtsov2018non, pino2019ergodic}.} \new{Moreover}, recent results in the MBL transition \cite{PhysRevB.99.094205} predict a typical localization length which follows exactly our Eq.~\eqref{qstarloc}. This strongly suggests that the Anderson transition on random graphs is in the same universality class as the MBL transition. Our approach could thus be used to characterize the critical behavior in this important problem.

\begin{acknowledgments}
We thank C.~Castellani, N.~Laflorencie, and T.~Thiery for fruitful discussions. We are grateful to G.~Parisi and F.~Ricci Tersenghi for mentioning the references \cite{morone2014large, PhysRevE.97.012152}. We acknowledge interesting discussions with K.S.~Tikhonov, A.D.~Mirlin and V.E.~Kravtsov concerning the critical properties.
We thank CalMiP for access to its supercomputer and the Consortium des \'Equipements de Calcul Intensif (C\'ECI).
This study has been supported through the EUR grant NanoX No. ANR-17-EURE-0009 in the framework of the ``Programme des Investissements d'Avenir'',  by  the  ANR  grant 
MANYLOK  (Grant No. ANR-18-CE30-0017), by  the  ANR  grant 
COCOA  (Grant No. ANR-17-CE30-0024-01), by the CONICET (Grant No. PIP 11220150100493CO), by ANCyPT (Grant No. PICT-2016-1056)
 and by the French-Argentinian LIA LICOQ.

\end{acknowledgments}

\bibliographystyle{apsrev4-1}
\bibliography{refs_SW.bib}

\end{document}